\documentclass[twocolumn,showpacs,preprintnumbers,amsmath,amssymb,graphicx]{revtex4-1}

\usepackage{graphicx}
\usepackage{dcolumn}
\usepackage{bm}

\begin{document}


\title{Revisiting the Wu-Yang Monopole: classical solutions and conformal invariance}

\author{J. A. O. Marinho}
\email{adnei@ita.br}
\author{O. Oliveira}
\email{orlando@teor.fis.uc.pt}
\affiliation{ CFC, Dep F\'{\i}sica, Universidade de Coimbra,   3004 516 Coimbra, Portugal}
\author{B. V. Carlson}
\email{brett@ita.br}
\author{T. Frederico}
\email{tobias@ita.br}
\affiliation{ Departamento de F\'{\i}sica, Instituto Tecnol\'ogico de Aeron\'autica, 12.228-900, S\~ao Jos\'e dos Campos, SP, Brazil}


\begin{abstract}
The Wu-Yang monopole for pure SU(2) Yang-Mills theory is revisited.  New classical solutions with finite energy are found for a generalized 
Wu-Yang configuration. Our method relies on known asymptotic series solutions and explores the conformal invariance of the theory to set
the scale for the monopole configurations. Furthermore, by breaking the scale invariance of the pure Yang-Mills theory by a term which
simulates the coupling to quark fields, four different monopole configurations with an energy of 2.9 GeV and spatial extent of 1.2 fm, 1.4 fm, 2.4 fm and
2.6 fm are obtained. These configurations may play a role in the transition to the plasma phase.
\end{abstract}


\maketitle

\section{Introducton and Motivation}

Classical solutions of Yang-Mills theory can
provide new insights into non-perturbative physics, giving hints
on infrared properties 
(related to confinement or chiral symmetry breaking, for example). 
Among the known classical solutions, we can name several configurations which have been
investigated and used in studies of hadronic phenomenology:
the instanton, meron, dyon and monopole (for a review, see \cite{actor}). 
Furthermore, in spite of the dyon being a kind of generalized monopole solution \cite{dyon}, and the existence of 
a possible meron-monopole connection \cite{cp}, 
the monopole solution seems to play a special role at the level of the basic structure of the theory.
Multipole solutions have also been discussed in the literature \cite{popov}.

Here we want to revisit the monopole solution, addressing the
generalized Wu-Yang monopole (WYM) \cite{WuYang}. 
The simple WYM is a Dirac magnetic monopole imbedded in the SU(2) gauge group, written in a
simple form as:
\begin{equation}
{A^0}^a=0~;~~~~~~{A^i}^a= \epsilon^{iab}\frac{x_b}{r^2}~.
\end{equation}
Among recent uses of the WYM, one could cite that the description 
of nuclear motion in a diatomic molecule can possess a Wu-Yang like-term \cite{MSW}. Moreover, the
screening effects in non-abelian Quark-Gluon Plasmas may give rise to Wu-Yang terms as
well \cite{QGP}. Furthermore, in \cite{stuler}  the monopole solution is used in the equation for the 
inverse of the quark propagator.
There, a spin-orbit term is computed together with an energy spectrum.

Less recent is the belief that a condensed phase of monopoles could encode the key of the confinement 
mechanism. Some authors argue that in the maximal abelian sub-algebra of the Yang-Mills theory, the monopole
assumes a crucial role in long distance effects, such that it gives rise to a so-called ``Abelian
dominance" \cite{AD}.

The WYM has also been used to investigate Gribov copies \cite{holdom}. This connects with the problem
of building a suitable generating functional for the Yang-Mills theory Green's functions 
valid beyond perturbation theory.

In this work, our goal is to obtain  solutions of the field equations assuming
a WYM configuration. With this in hands, one can estimate an energy scale and study Gribov copies, 
among other applications. 

The classical configurations that we are going to discuss are of the generalized WYM type
\begin{equation}
{A^i}^a= \epsilon^{iab}\frac{x_b}{r^2}f(r)~.\label{potential}
\end{equation}
Of course the generalized WYM keeps its primitive nature. It mixes the color and the space indices in the same 
way, and includes the original WYM as a particular case. The corresponding field equation is a nonlinear
ordinary differential equation for $f(r)$
that will be investigated with a combination of analytical and numerical tools.
This paper is organized as follows. In section \ref{GWYmonopole}, the generalized Wu-Yang configuration
is discussed, with the computation of $F^a_{\mu\nu}$ and the classical energy. In  \ref{DSolSeries} the solutions
of the classical equation of motion as power series is discussed. In section \ref{DSolNum} two different numerical solutions of
the classical field equation are computed and their chromomagnetic fields, for large and small $r$, are discussed.
In section \ref{sec_energy} we explore the conformal invariance of the theory to introduce a mass scale of the order of
a typical nonperturbative scale $\sim 1$ GeV (or $\sim 1$ fm) for the monopole. Then, one can estimate the monopole size (energy).
Monopole like configurations with discontinuous first derivatives
are discussed in \ref{SolDescontinuas}. Finally, in  \ref{resultados} we resume and conclude.

\section{Generalized Wu-Yang Monopole Ansatz \label{GWYmonopole}}

In this work, we are going to treat several classical aspects of the
Yang-Mills theory, given by the Lagrangian density $\mathcal{L} \,
= \, - \frac{1}{4} \, F_{\mu\nu}^a \, F^{ \, \mu\nu}_a$. From now
on, it will be assumed that the rescaling of the gauge fields has
been performed, so that the coupling constant $g$ will be omitted.
Furthermore, only the SU(2) gauge group will be
considered, for the sake of simplicity.

The classical equations of motion are a set of coupled non-linear
partial differential equations. Among the possible classical
configurations, one may take the generalized Wu-Yang monopole discussed previously. The time
component of the gauge field vanishes, $A_0^a=0$, and the spatial
components are given by equation (\ref{potential}). Note that this static configuration satisfies the 
Lorentz, $\partial^\mu A^a_\mu = 0$, and Coulomb , $\nabla \cdot \vec{A}^a = 0$, gauge fixing conditions
 automatically.

Here one might of course pose the question of the completeness of
this configuration. Indeed, the $f(r)$ function accounts for 
distance  fluctuations. However, in terms of directions in
the color space, one could think that the monopole configuration
is very restricted. Further, we see that the mixing of color and
space-time indices is also a quite strong restriction. 

Let us go ahead and calculate the Lagrangian density. It is just a
matter of algebraic work to find the form of the nonabelian gauge
tensor due to the generalized Wu-Yang configuration. It reads
\begin{eqnarray}
{F^a_{ij}}& = &
 \left( \frac{f^\prime}{r^3} - 2 \frac{f}{r^4} \right)\left[ x^i \epsilon_{jab}x^b ~ -  ~ x^j\epsilon_{iab} x^b  \right]
   \nonumber \\
   & &   ~  + ~ 2 \frac{f}{r^2} \, \epsilon_{ija} ~ - ~  \frac{f^2}{r^4} \, \epsilon_{ijb} x^b \, x^a \, ,  \label{chromofields} \\
  F^a_{0j} & = & 0 \, ;   
\end{eqnarray}
Note that the chromoelectric field vanish. Another important object is the energy density, obtained by the
contraction of the $F_{\mu\nu}$ tensor. Then, the corresponding chromomagnetic energy density is
\begin{equation}
\frac{1}{4} (F^a_{ij})^2  =  \left( \frac{f^\prime}{r} \right)^2 +
2 \, \frac{f^2}{r^4}  -  2 \, \frac{f^3}{r^4}
       +  \frac{1}{2} \frac{f^4}{r^4} ~.
       \label{eq_000}
\end{equation}
The total energy is given after integrating (\ref{eq_000}) over the space volume
Now, it is convenient  to make the replacement $f = g + 1$, which gives the energy as:
\begin{equation}
\mathcal{E}  =  4 \pi  \int^{+ \infty}_{0} dr ~
   \left\{ \left( g^\prime \right)^2   ~ + ~
             \frac{\left( g^2 - 1 \right)^2}{2 \, r^2} \right\}  .
\label{Ewuyang}
\end{equation}
Of course, depending on the form of the integrand, $g(r)$ may have 
a confining shape or not. Note that the requirement of finite energy prevents $g(r)$ of growing without bound with
$r$.
To achieve the solutions for
$g(r)$, one has to solve the equations of motion, obtained by
requiring that $g(r)$ minimizes the energy. The equation of motion reads:
\begin{equation}
  g^{\prime\prime} ~ = ~ \frac{g}{r^2} \left( g^2 - 1 \right) \,   .
  \label{feq1}
\end{equation}
This equation, obtained by minimizing the energy, is exactly the same one that is derived 
using (\ref{potential}) in the classical equations of  motion of the classical Yang-Mills theory.
Of course, first one has to solve equation (\ref{feq1}) 
to compute the classical energy from the generalized WYM configurations.

\section{Series Solution \label{DSolSeries}}

Given that (\ref{feq1}) is nonlinear, it is quite difficult if not impossible to get an analytical solution. 
Then, let us construct a series to approximate the solution. 
First note that equation (\ref{feq1}) is symmetric under scale 
transformations: if $g(r)$ is a solution, then $g(\lambda r)$ is also a solution, for any $\lambda>0$. To
see this, multiply the equation by $\lambda^{2}$ and convert it to an identical equation in the new variable 
$x=r/\lambda$, whose solution is $g(\lambda x)$. Next, note that any finite solution must remain in the 
interval $-1\le g(r)\le1$. If $g(r)>1$ with $g^{\prime}(r)>0$, a necessary condition for approaching the value $1$
from below, the solution grows without bound. If $g(r)< -1$ with $g^{\prime}(r)<0$, 
a necessary condition for approaching $-1$ from above, the solution diminishes 
without bound. Furthermore, for a solution of the equation to be finite in the limits of its domain,
we must have
\begin{equation}
g(r) \rightarrow\pm\left(1-\left(\lambda_{l}r\right)^{2}\right)\qquad
r\rightarrow0
\end{equation}
and
\begin{equation}
g(r)\rightarrow\pm\left(1-\frac{1}{\lambda_{h}r}\right)\qquad
r\rightarrow\infty\,,
\end{equation}
for some $\lambda_{l}$ and $\lambda_{h}$. Further, we note that if $g(r)$
is a solution, then $-g(r)$ is as well.

This is sufficient to give us a hint about the necessary form of
the required series solution. We cast the solution $g(r)$ as:
\begin{equation}
  g(r) ~ = ~ \left\{ \begin{array}{lll}
                    A(r)=\sum_{n=0}\limits^{N_A} a_n \, r^n, &  $\hspace{0.2cm}$  & (0 \leq r\leq R) \\
                     & & \\
                    B(r)=\sum_{n=0}\limits^{N_B} b_n \, r^{-n}, & $ \hspace{0.2cm}$  & (R \leq r )
                    \end{array} \right.
                    \label{gseries}
\end{equation}
The reader should be aware that the first terms of these series are well known -
see, for instance, appendix G of \cite{actor}.

The function $g(r)$ presents distinct behaviors depending
on the distance to the origin. For small r, $A(r)$ favors growth of the
interaction with increasing distance. For r greater than R,
$B(r)$ favors the decrease of the interaction with distance.

\begin{figure}[t]
   \centering
   \includegraphics[scale=0.35]{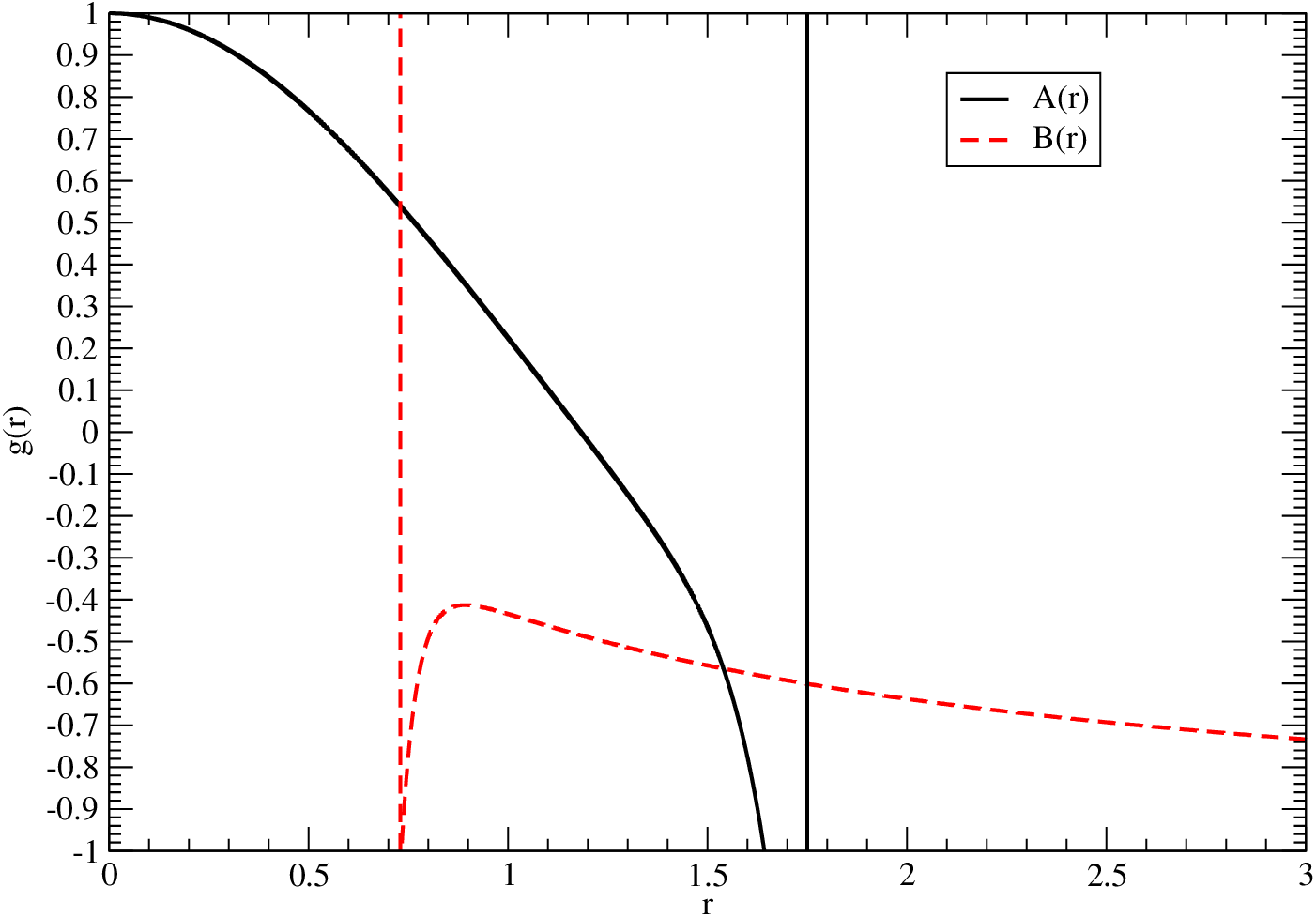} 
   \caption{The series solutions $A(r)$ and $B(r)$. The vertical lines are the naive radius of convergence for each series - see text for details.
                  \label{fig_solseries}}
\end{figure}

Using the above solution in the differential equation (\ref{feq1}), a recursion relation for the various
$a_n$ and another for the $b_n$ coefficients can be derived. Indeed, to lowest order in $r$  or in $1/r$ for the
$A(r)$ and $B(r)$ functions, respectively, one has that $a_0^3-a_0=b_0^3-b_0=0$. Therefore
 $a_0=0,\pm 1$ and $b_0=0,\pm 1$. The zero value implies always the trivial solution, i.e. $a_n = 0$ and/or
 $b_n=0$. Given the symmetry $g(r) \rightarrow - g(r)$, we will consider only the $+1$ solution. Furthermore,
 for $A(r)$ it turns out that the coefficients for odd powers of $r$
 vanish. After redefining the $A(r)$ series to include only even powers
 of $r$ 
and using the $+1$ value for the $a_0$ and $b_0$, the $a_1$ and $b_1$ coefficients are arbitrary and the
remaining coefficients are given in terms of $a_1$ and $b_1$.

In the following, in the computation of
$A(r)$ and $B(r)$ the truncations $N_A = N_B = 18$ are used everywhere. For  completeness, we write the first terms of
the series solutions. For $a_1 = b_1 = 1$, it comes that
\begin{eqnarray}
  A(r) & = &  1 - r^2   + \frac{3}{10} r^4 - \frac{1}{10} r^6 + \frac{59}{1800} r^8 - \frac{71}{6600} r^{10}  \nonumber \\
         &     &
                    + \frac{15143}{4290000} r^{12} - \frac{20327}{17550000} r^{14} + \frac{1995599}{5250960000} r^{16} \nonumber \\
         &     &
                    - \frac{311031533}{2494206000000} r^{18} 
                    \label{serie_A18}
\end{eqnarray}
and
\begin{eqnarray}
  B(r) & = &  1 - \frac{1}{r} + \frac{3}{4} \frac{1}{r^2} - \frac{11}{20} \frac{1}{r^3}  + \frac{193}{480}\frac{1}{r^4} - \frac{47}{160}\frac{1}{r^5} \nonumber \\
         &    &
         + \frac{3433}{16000}\frac{1}{r^6} - \frac{67699}{432000}\frac{1}{r^7} + \frac{1318507}{11520000}\frac{1}{r^8} \nonumber \\
         &     &
         - \frac{2118509}{25344000}\frac{1}{r^9} \, .
          \label{serie_B9}.
%
%
\end{eqnarray}
We call the reader attention that, due to conformal invariance, $A(r)$ and $A( \lambda r)$ or $B(r)$ and $B( \lambda r)$ are solutions of the equations
of motion.  For the truncation described above, one can estimate the radius of convergence of the series from the higher order terms in the series.
For $A(r)$ convergence occurs for $r < 1.75$, while for the $B(r)$ series convergence happens for $r > 0.73$.

We have investigated how the naive estimate of the radius of convergence changes with the order of the truncation, i.e. with $N_A$
and with $N_B$. The analysis of the series for $A(r)$ and $B(r)$ show that, with the exception of the first few terms, the estimate of the radius of 
convergence is almost independent of $N_A$ and $N_B$. We can then consider matching the series $A(r)$ to the value and first derivative of the 
series $B(\lambda r)$ for some $\lambda$ and some $r$ in the interval $0.73/ \lambda < r < 1.75$. It can be shown, however, that such values of 
$\lambda$ and $r$ do not exist, which prevents the construction of a global solution of (\ref{feq1}) in the form of the power series.


The series solutions can be seen in figure \ref{fig_solseries} together with the naive radius of convergence.

\section{Numerical Solutions \label{DSolNum}}

A numerical solution of equation (\ref{feq1}) can be built combining the series for $A(r)$ and $B(r)$, described in the previous section, with a Taylor
expansion of $g(r)$. If at $r = r_0$  the solution $g(r)$ along with its first and second derivatives are known, one can expand $g(r)$ around $r_0$
\begin{equation}
   g(r_0 + \delta r) = g(r_0) + g'(r_0) \delta r  +  \frac{1}{2} g''(r_0) ( \delta r)^2 + \cdots
\end{equation}
and similar expansions for $g'(r)$ and $g''(r)$ can be written. Replacing, in the differential equation (\ref{feq1}), 
the expansion for $g(r_0 + \delta r)$ and $g''(r_0 + \delta r)$, working order by order in $\delta r$, one can compute the 
derivatives $g^{(n)}(r_0)$ for $n = 3, \, 4, \,\dots$ in terms of $g(r_0)$, $g'(r_0)$ and $g''(r_0)$. In our numerical solutions, discussed below, we 
have considered expansions of $g(r)$ up to order 6.

For the solution called $g_{int}(r)$, the integration starts at $r=0$, setting $g(r) = A(r)$, and using an integration step of $\delta r = 10^{-4}$.
The series solution was used up to the point where the relative error between the l.h.s. and r.h.s of equation (\ref{feq1}) was below a certain error,
which we toke to be $10^{-6}$. If the solution $g \sim 0$, then instead of the relative error we have used the absolute error,
i.e. the absolute difference between the l.h.s. and r.h.s. of the differential equation. 
This procedure (so far, using A(r)) gives $g(r)$, $g'(r)$ and $g''(r)$, as the series for $A(r)$ up to $r = 0.61$. From this point onwards, $g(r)$ was propagated
using the Taylor series procedure described at the beginning of this section. At each step the relative error, or the absolute error, in the differential
equation was checked. If the precision was above the tolerance demanded, then the propagation was restarted using the last point where the
differential equation was satisfied to calculate the coefficients of the Taylor expansion. The procedure was repeated up to the point
where $|g(r)| = 1$. From this point onwards it was assumed that $g(r) = \pm 1$ for all $r$, with the sign chosen to produce a continuous function.

For the solution called $g_{ext}(r)$, integration was started at $r = 300$, setting $g(r) = - B(r)$, and using $\delta r = - 10^{-4}$ as integration step. 
The procedure to compute $g_{ext}(r)$ follows the same lines as the computation of $g_{int}(r)$, with the difference that the solution is now 
propagated backwards. Deviations from the $B(r)$ series solution happen for $ r \le 2.12$.

The numerical integration was checked exploring different integration steps and the conformal invariance of equation (\ref{feq1}). 
In particular, the scaling of $g'(r)$ and $g''(r)$ under a scale transformation was checked carefully. Furthermore, relaxing the precision on the
differential equation, we observed that the numerical solution follow the series solution almost up to its naive convergence radius.
The several tests performed gave us confidence on our numerics.

\begin{figure}[t]
   \centering
   \includegraphics[scale=0.37]{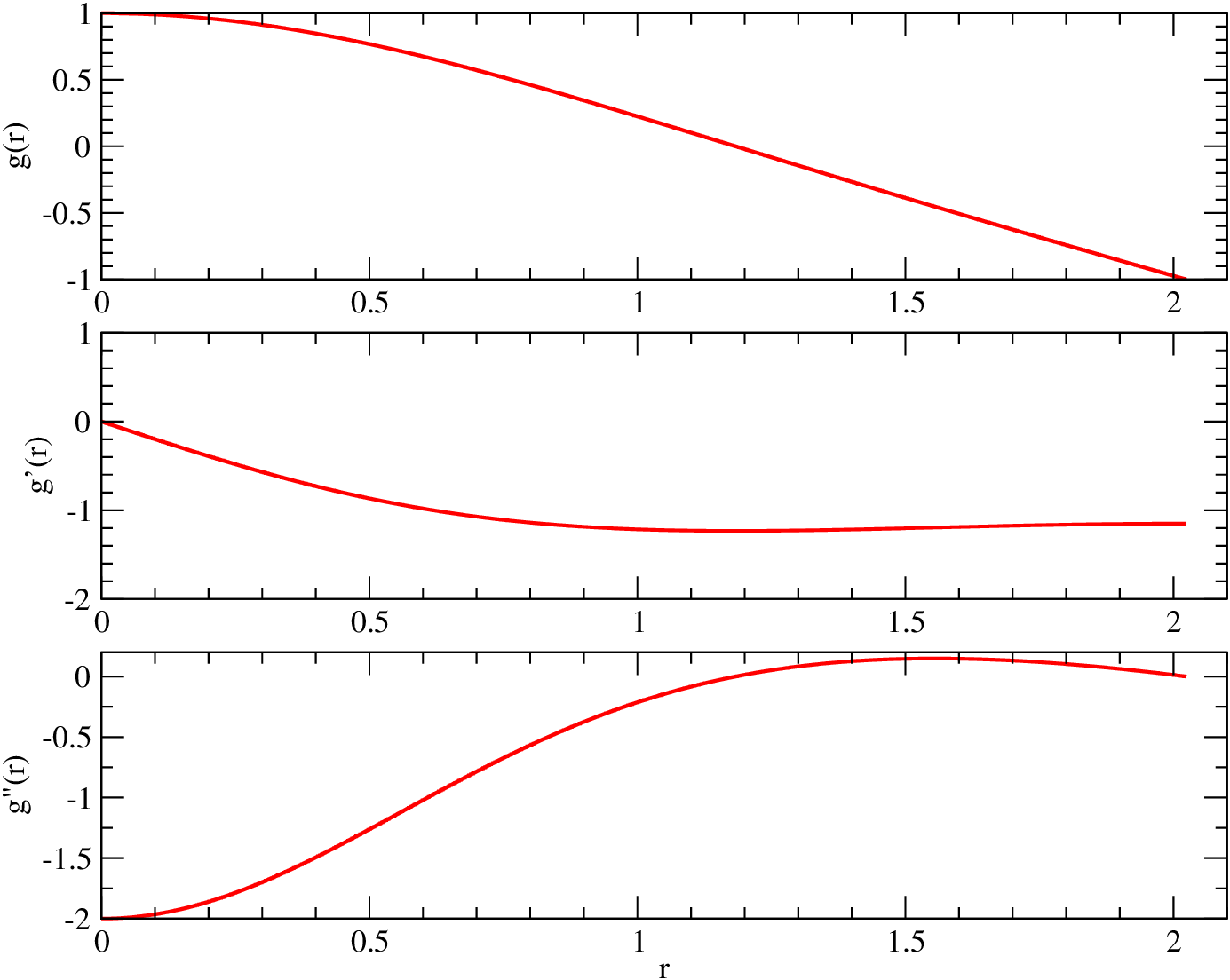} 
   \caption{The solution $g_{int}(r)$ and its derivatives.  \label{fig_sol_gint}}
\end{figure}

\begin{figure}[t]
\vspace{0.5cm}
   \centering
   \includegraphics[scale=0.37]{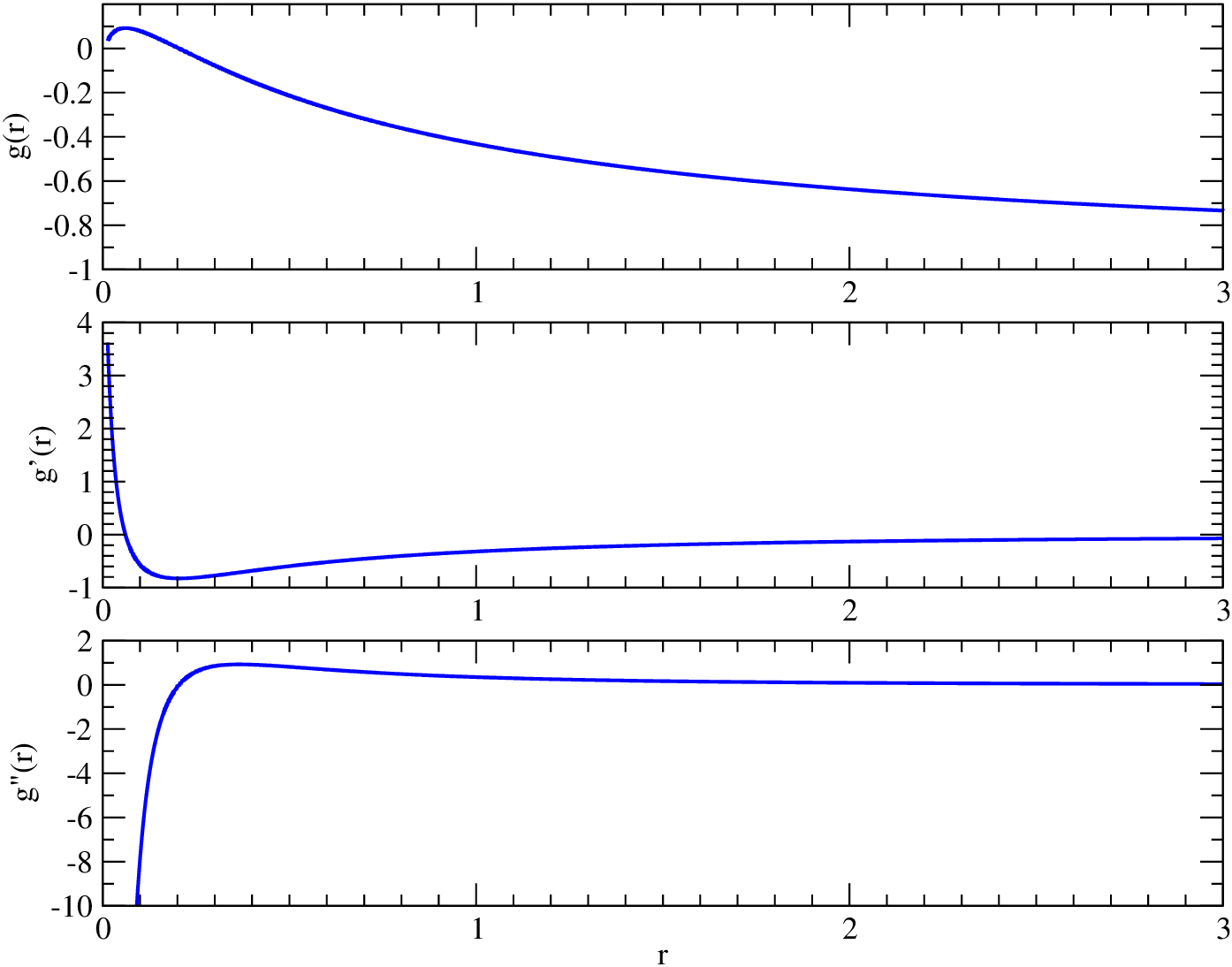} 
   \caption{The solution $g_{ext}(r)$ and its derivatives.   \label{fig_sol_gext}}
\end{figure}

The curves for $g_{int}(r)$ and $g_{ext}(r)$, including its first and second derivatives, are shown in figures \ref{fig_sol_gint} and \ref{fig_sol_gext},
respectively.

 Let us discuss first the solution $g_{ext} (r)$. $g_{ext} (r)$ is the numerical solution found by Wu and Yang in \cite{WuYang}. Although in figure
 \ref{fig_sol_gext}, it seems that the curve touches $r=0$, this is not the case. A closer look shows that the smallest value of $r$ numerically available is 0.0133. Our numerical procedure was not able to go below this value of $r$. Anyway $g_{ext} (r)$ matches perfectly the solution of the linearized field
equation (\ref{feq1}), i.e.
\begin{equation}
 g(r) = \sqrt{r}  \left( G_0 \sin \left( \frac{\sqrt{3}}{2}\ln r \right) + G_1 \cos \left( \frac{\sqrt{3}}{2}\ln r \right) \right),
    \label{WuYangNum_r0}
 \end{equation}   
with $G_0 = -0.079223$ and $G_1 = -0.4251983$. $g_{ext} (r)$ oscillates as $r$ approaches zero and its derivative diverges at $r = 0$.
The classical energy, as given by (\ref{Ewuyang}), associated with $g_{ext}(r)$ is infinite.

\begin{figure}[t]
   \centering
   \includegraphics[scale=0.37]{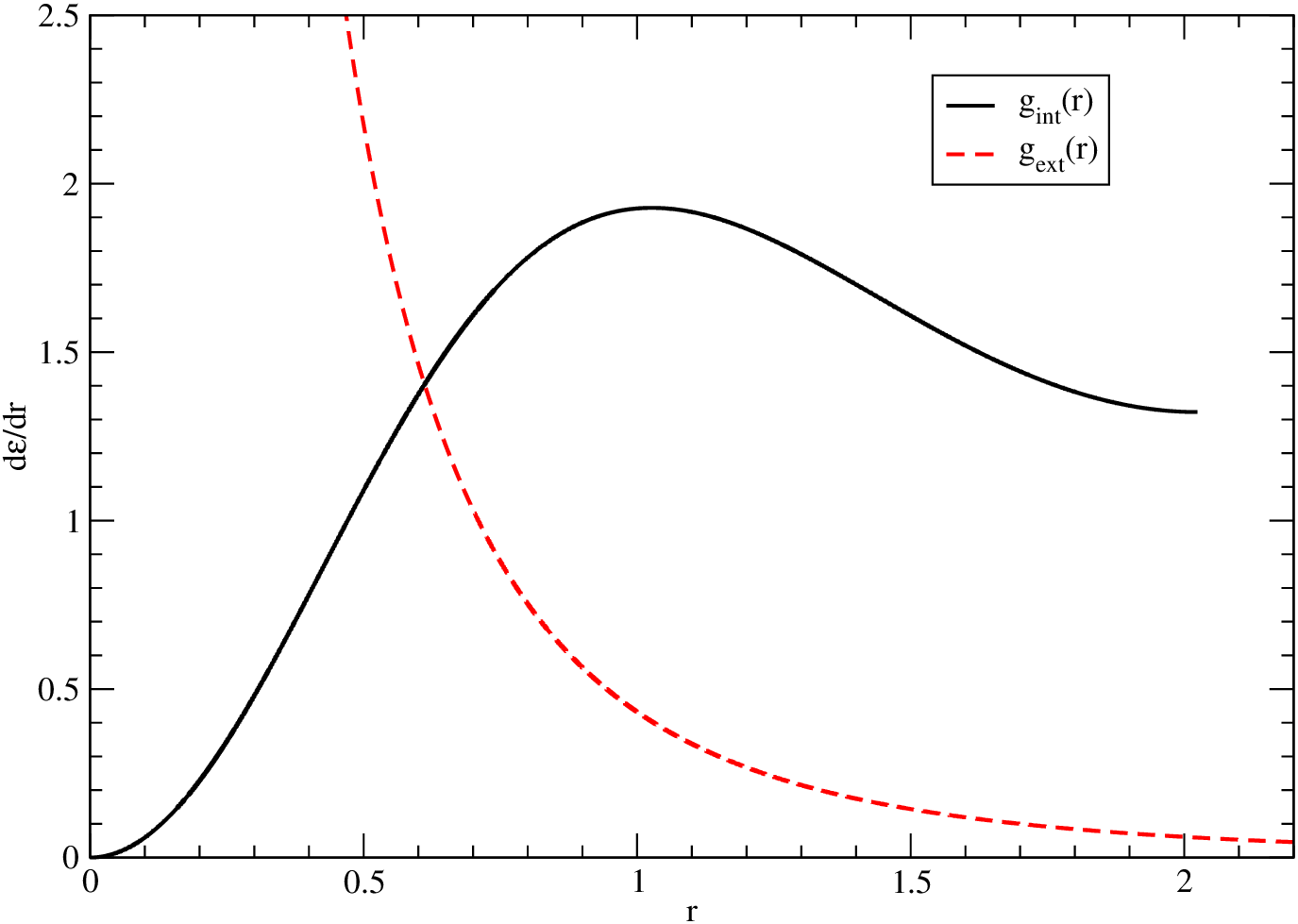} 
   \caption{The energy density for $g_{int}(r)$ and $g_{ext}(r)$.  \label{fig_e_density}}
\end{figure}

On the other hand, the solution $g_{int} (r)$ has finite energy, $\mathcal{E} = 33.63$ in dimensionless units, but its first derivative 
is discontinuous at $r = 2.024$ where
$g = -1$. 
For $r > 2.024$ one assumes a constant solution with $g = -1$ and for $r > 2.024$ the energy density associated with $g_{int} (r)$
vanishes.
The energy density for $g_{int}(r)$ and $g_{ext}(r)$ is reported in figure \ref{fig_e_density}.

\subsection{The Chromomagnetic Field}

The chromomagnetic fields are given in terms of $f = 1 + g$ by equation (\ref{chromofields}). For each of the numerical solutions described previously,
one can build two different $f = 1 \pm g$ and its large distance behavior is given by
\begin{equation}
  g_{int}(r)  =  -1 \qquad\mbox{and}\qquad   g_{ext}(r) = 1 - \frac{1}{r} \, ,
\end{equation}
i.e.
\begin{equation}
  f(r)  =  \left\{ \begin{array}{lll}
                    0, & & g = + g_{int} \\
                       & & \\
                    2, & & g = - g_{int} 
                    \end{array}\right.
\end{equation}
and
\begin{equation}
  f(r) = \left\{ \begin{array}{lll}
                  2 - \frac{1}{r} , & & g = + g_{ext} \\
                  & & \\
                  \frac{1}{r} , & & g = - g_{ext} \, .
                                      \end{array}\right.
\end{equation}
For the first solution $F^a_{ij} = 0$. For the second and third solutions, $F^a_{ij} \sim 4/r^2$ at large $r$. Finally, for the last solution the 
chromomagnetic field associated with monopole vanishes faster and goes like $F^a_{ij} \sim 1 / r^3$ at large distances. Given the behavior of 
$F^a_{ij}$ at large distance, one can say that solutions two and three are monopole like solutions, i.e. $F^a_{ij} \sim 1/r^2$, 
while the fourth solution is a dipole like solution, i.e. $F^a_{ij} \sim 1/r^3$.

In what concerns the short distance behavior, for all configurations $F^a_{ij}$ diverges at $r = 0$. 
The exception being the trivial solution with a vanishing chromomagnetic field.

\section{Wu-Yang Monopoles and Classical Energy \label{sec_energy}}

In the previous section we have computed monopole configurations with infinite classical energy (those associated with $g_{ext}(r)$), and
monopoles with finite energy (associated with $g_{int}(r)$). However, in what concern monopoles with finite energy, besides the solutions associated with
$g_{int}(r)$, the constant solutions of the equation of motion (\ref{feq1}) also belong to this family of functions.
The equation of motion (\ref{feq1}) has the constant solutions $ g = 0, \pm 1$ but only $g = 1$ and $g = -1$ have finite zero energy - see the definition 
(\ref{Ewuyang}). In particular, $g = -1$ is associated with the trivial $A^a_\mu = 0$ configuration and will not be considered any further. 

The monopole with $g = 0$, i.e. $f = 1$, despite having infinite classical energy, has an
associated chromomagnetic field vanishing as $1/r^2$ for large $r$.

For the monopole configurations with classical infinite energy, the divergence in $\mathcal{E}$ is associated with an ultraviolet, i.e. short distance,
divergence of the energy density. Like in classical electrodynamics, one can cure the divergence via the introduction of a short distance
cutoff. Then, a scale is introduced into the theory and conformal invariance is broken explicitly. 

On the other hand, finite energy monopoles are either constant solutions of (\ref{feq1}) or have a discontinuity on the first derivative of $f$.

Let us go back to the solution $g_{int}(r)$. Conformal invariance means that if $g_{int}(r)$ is a solution of equation (\ref{feq1}), 
$g_{int}( \lambda r)$ is also a solution of the same differential equation with energy $33.63 / \lambda$. 
Assuming that this energy is a typical hadronic scale
$\sim 1$ GeV, then $ \lambda = 33.63$ GeV$^{-1}$ and the discontinuity occurs for $r_d = 2.024 \, \lambda = 13.4$ fm, roughly one order of magnitude
larger than the typical hadronic scale $\sim 1$ fm. If instead, one uses the mass of the lowest glueball to set the scale $M_{glue} = 1.7$ GeV \cite{chen},
these figures should be divided by 1.7, i.e. $\lambda =  19.78$ GeV$^{-1}$ and $r_d = 7.9$ fm. Again, the discontinuity on the first derivative of $f$ 
happens at a rather large length scale.

The hadronic length scale $\sim 1$ fm comes from the analysis of baryons and mesons constituted by quarks. 
Our reasoning shows that if the finite energy solution built from $g_{int}(r)$ plays a role on the dynamics of the Yang-Mills theory, then the large distance 
behavior is clearly dominated by gluon fields. Furthermore, these gluon fields have a spatial size about ten times larger than the usual quark content. 
Ignoring, for the moment, possible implications of such a picture on quark matter phenomenology, in the next section
we investigate how far can one take the discontinuous monopole configurations.

\section{Discontinuous Wu-Yang Monopoles With Finite Energy \label{SolDescontinuas}}

Discontinuities in the derivative can be introduced in the field equation (\ref{feq1}) via a Dirac-delta function $\delta(r - R_0)$.  
Away from $r =R_0$ one recovers the original equation and $g(r)$ is given by one of the solutions discussed in section \ref{GWYmonopole}. 
Adding a term like $\zeta \, \delta(r - R_0)$, where $\zeta$ is a constant, to (\ref{feq1}) is equivalent to add $\zeta \, g(r) \delta(r - R_0)$ to the
density of energy. Remember that the field equation was obtained directly from the classical Yang-Mills equations for the generalized Wu-Yang
configuration and from minimizing the energy (\ref{Ewuyang}). Such a term can be due to the presence of static quarks located at $R_0$. In this
sense one can see $\zeta$ as the square of the quark wave function at $R_0$ and $R_0$ as a mean value of the quark radius.
Furthermore, the introduction of the new term in $\mathcal{E}$ for a fixed $R_0$ breaks conformal invariance.

A physical scale should be introduced in QCD in order to construct the observables. In hadronic physics this is set by $\Lambda_{\text{QCD}}\sim 1\; {\text fm}^{-1}$, i.e.,  the scale where conformal invariance is broken, as in modern bottom-up approaches of the hadronic spectrum  within gauge/gravity duality (see e.g. \cite{brodsky}). However, our classical solutions, as we will see, attain too large energies, in the hadronic scale, when the typical extension of the chromagnetic field is of the order of 1 fm, which suggests that it is required a collective phenomena to excite such field configurations. More on that  discussion will come later on.

\begin{figure}[t]
   \centering
   \includegraphics[scale=0.35]{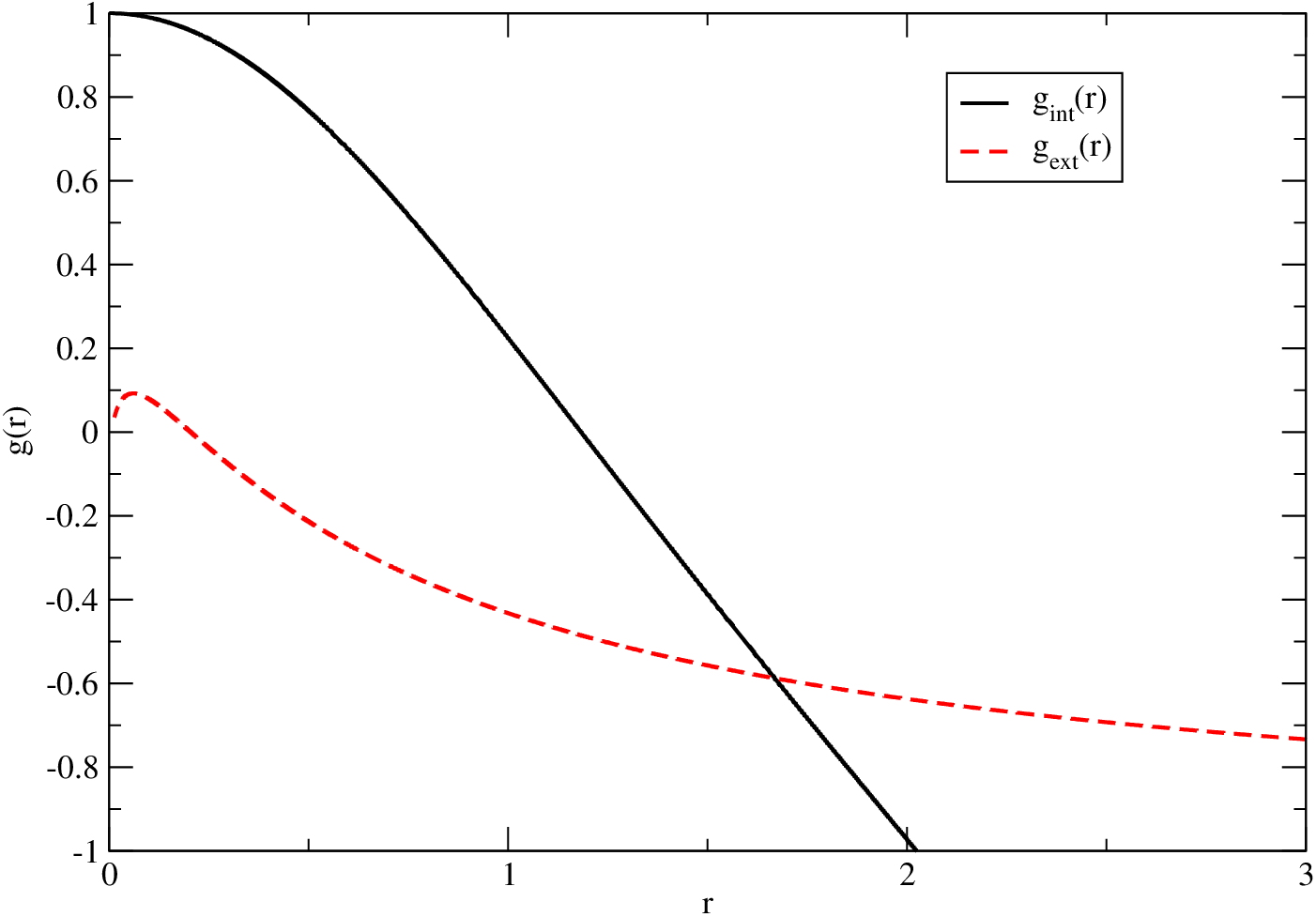} 
   \caption{The solutions $g_{int}(r)$ and $g_{ext}(r)$.  \label{fig_gint_gext}}
\end{figure}

So let us assume that the classical field equation is given by
\begin{equation}
  g^{\prime\prime} ~ = ~ \frac{g}{r^2} \left( g^2 - 1 \right) \,   + \, \zeta \, \delta(r - R_0).
  \label{eq_new}
\end{equation}
For $r \ne R_0$ the solutions of (\ref{eq_new}) are the functions $g(r)$ discussed in section \ref{GWYmonopole} - see figure \ref{fig_gint_gext}.
Assuming that $g(r)$ is continuous at $R_0$, integrating (\ref{eq_new}) from $R_0 - \epsilon$ to $R_0 + \epsilon$ gives, in the limit where $\epsilon \rightarrow 0^+$,
\begin{equation}
  g' ( R_0 + \epsilon ) - g' ( R_0 - \epsilon ) =  \zeta \, .
  \label{eq_new1}
\end{equation}

\begin{figure}[t]
\vspace{0.6cm}
   \centering
   \includegraphics[scale=0.35]{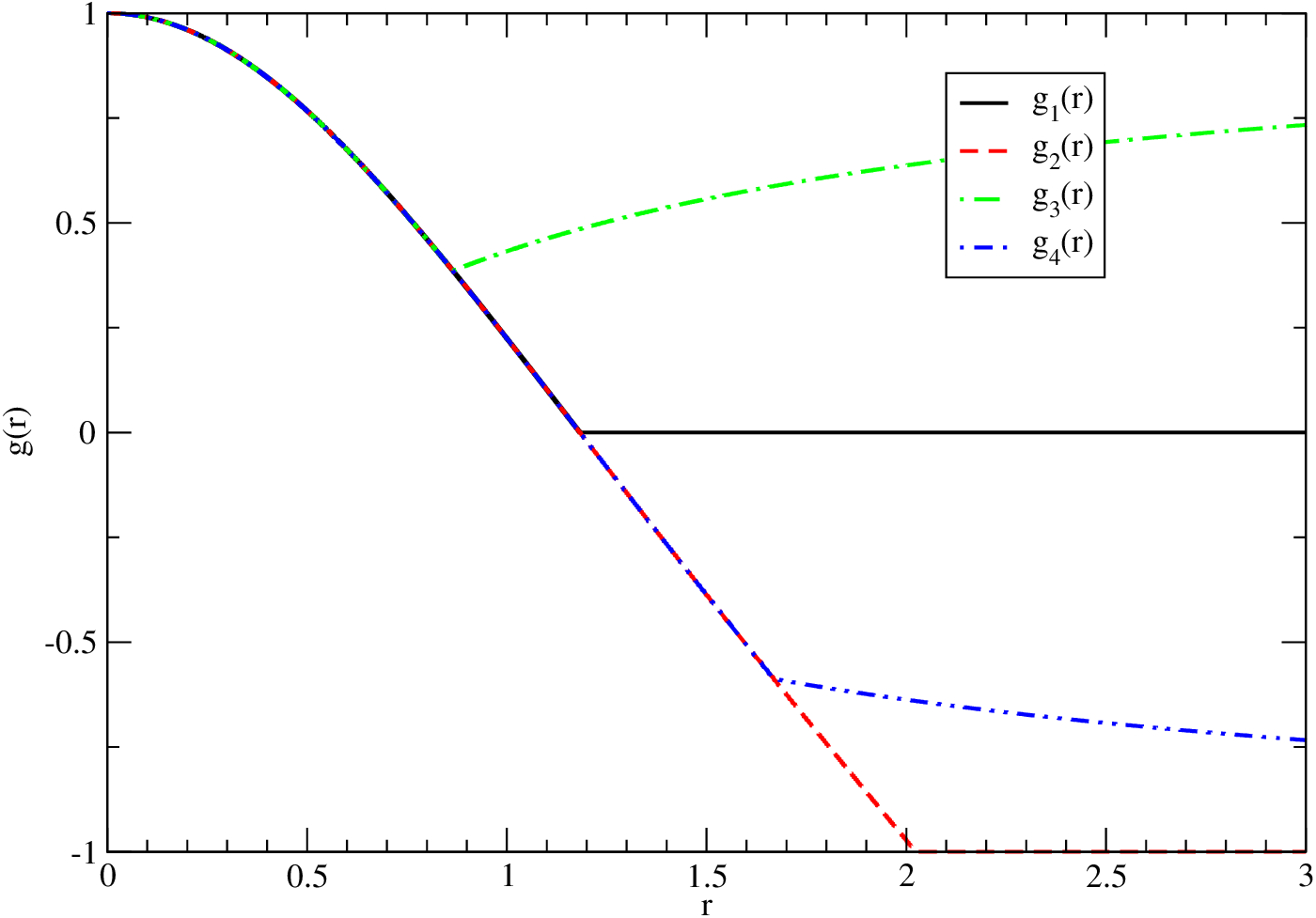} 
   \caption{The solutions $g_1(r)$ to $g_4(r)$.  \label{fig_g_1_2_3_4}}
\end{figure}

Finite energy solutions require $g(0) = \pm 1$. Therefore, for $r < R_0$ only $\pm g_{int} (r)$ and the constant solutions $g = \pm 1$ are acceptable.
On the other side, for $r > R_0$, there is no reason to exclude any of the solutions $\pm g_{int} (r)$, 
$\pm g_{ext} (r)$, $g(r) = 0$ or $g(r) = \pm 1$ which can be combine to produce several finite energy solutions. 
The solution of equation (\ref{eq_new}) gives an $f$ continuous and $f^\prime$ discontinuous at $R_0$ -- see equation (\ref{eq_new1}).
The various solutions are
\begin{equation}
    g_1(r) = \left\{ \begin{array}{ccl}
                        g_{int}(r)  &  & r \le R_0 = 1.1831 \\
                        &  &  \\
                        0              &  & r > R_0
                       \end{array} \right.
                       \label{monopolod1}
\end{equation}
for a $\zeta_1 = 1.2327$,
\begin{equation}
    g_2(r) = \left\{ \begin{array}{ccl}
                        g_{int}(r)  &  & r \le R_0 = 2.0240 \\
                        &  &  \\
                        -1              &  & r > R_0
                       \end{array} \right.
                       \label{monopolod2}
\end{equation}
with a $\zeta_3 = 1.1498$,
\begin{equation}
    g_3(r) = \left\{ \begin{array}{ccl}
                        g_{int}(r)     &  & r \le R_0 = 0.8651 \\
                        &  &  \\
                        -g_{ext}(r)     &  & r > R_0
                       \end{array} \right.
                       \label{monopolod3}
\end{equation}
with a $\zeta_3 = 1.5432$ and
\begin{equation}
    g_4(r) = \left\{ \begin{array}{ccl}
                        g_{int}(r)  &  & r \le R_0 = 1.6685 \\
                        &  &  \\
                        g_{ext}(r) &  & r > R_0
                       \end{array} \right.
                       \label{monopolod4}
\end{equation}
with $\zeta_2 = 1.009$. In all cases $\zeta$ was computed with equation (\ref{eq_new1}). The various solutions can be seen in figure
\ref{fig_g_1_2_3_4}.

Assuming that the energy density is modified by the term $\zeta \, g(r) \, \delta(r - R_0)$, then the total energy is given by
\begin{equation}
\mathcal{E}   =  4 \pi \left(  \int^{+ \infty}_{0} dr ~
   \left\{ \left( g^\prime \right)^2   ~ + ~
             \frac{\left( g^2 - 1 \right)^2}{2 \, r^2} \right\}  ~ + ~ \zeta \, g(R_0) \right) .
\end{equation}
Since $g_1 (R_0) = 0$, $g_2(R_0) = -1$, $g_3(R_0) = 0.3859$ and $g_4(R_0) = -0.5877$, the energy of the monopoles is, in dimensionless units,
$\tilde{\mathcal{E}}_1 = 17.28$, $\tilde{\mathcal{E}}_2 = 19.18$, $\tilde{\mathcal{E}}_3 = 20.90$, $\tilde{\mathcal{E}}_4 = 21.06$, respectively. 
In physical units $r = \lambda \tilde{r}$, 
$\mathcal{E} = \tilde{\mathcal{E}} / \lambda$, where the quantities with tilde are dimensionless. Setting the energy of the monopole to 1 GeV, then
$\lambda_1 = 17.28$ GeV$^{-1} = 3.410$ fm,
$\lambda_2 = 19.18$ GeV$^{-1} = 3.785$ fm,
$\lambda_3 = 20.90$ GeV$^{-1} = 4.124$ fm,
$\lambda_4 = 21.06$ GeV$^{-1} = 4.156$ fm. 

In physical units, the discontinuity in the derivative happens at
$\lambda \, R_0$. For $g_1(r)$, this means $4.0$ fm. For $g_2(r)$ one gets $\lambda \, R_0 = 7.7$ fm.
For $g_3(r)$ one gets $\lambda \, R_0 = 3.6$ fm
and for $g_4(r)$ it follows that
$\lambda \, R_0 = 6.9$ fm. 
Note that if one uses the lowest glueball mass 1.7 GeV \cite{chen} to set the scale, one should divide these figures by 1.7, i.e. the discontinuities in
the derivative would occur at 2.4 fm, 4.5 fm, 2.1 fm, 4.1 fm for $g_1(r)$, $g_2(r)$, $g_3(r)$. $g_4(r)$, respectively, beyond which the chromomagnetic fields would decay,
at least, as $1/r^2$. Again, the scale at which $r_d = R_0$ happens is larger, at least two times, than the typical hadronic scale.

\begin{figure}[t]
   \centering
   \includegraphics[scale=0.35]{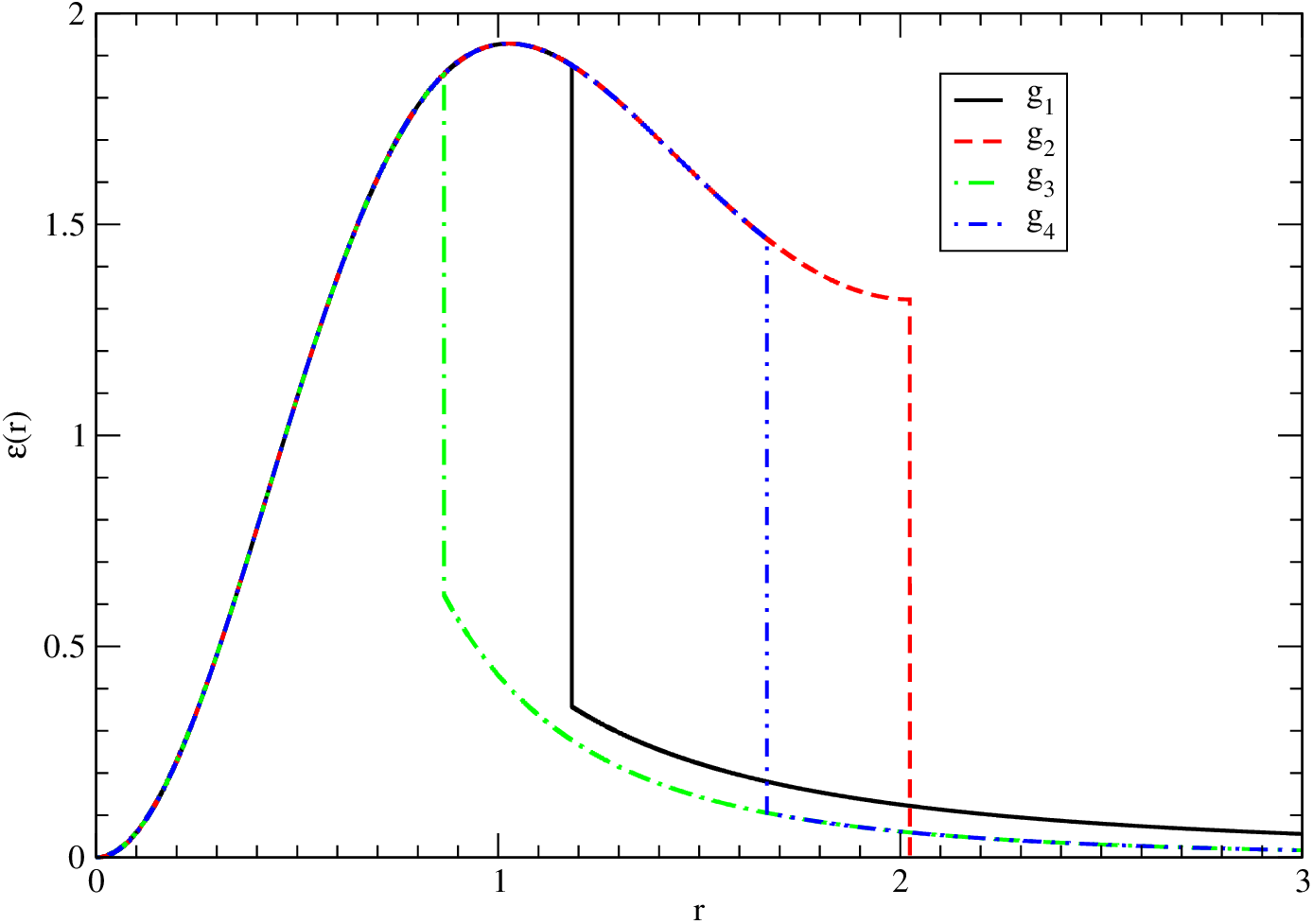} 
   \caption{Energy density, in dimensionless units, for the finite energy solutions $g_1(r)$, $g_2(r)$ and $g_3(r)$. Note the $1/r^2$ dependence at large
                $r$ for $g_1(r)$, $g_3(r)$ and $g_4(r)$. \label{fig_dens_finite_e}}
\end{figure}

In dimensionless units, the energy density for the various solutions under discussion is reported in figure \ref{fig_dens_finite_e}. 
Another measure of the monopole size is given by the maximum in the energy density. For $g_1(r)$, $g_2(r)$ and $g_4(r)$ 
the maximum occur at $r = 1$. For $g_3(r)$ the maximum occur at $r = 0.8651$. 
Given the $\lambda$ values computed in the previous paragraph, if one assumes a scale of 1 GeV, then the monopole sizes are 
3.4 fm,  4.5 fm,  3.6fm and 4.2 fm for $g_1(r)$, $g_2(r)$, $g_3(r)$  and $g_4(r)$, respectively. 
Otherwise  if one uses the glueball mass to set the scale, the characteristic sizes are
2.0 fm, 2.2 fm, 2.1 fm and 2.4 fm, respectively.

\section{Results and Discussion  \label{resultados}} 

In this work  we have revisited the Wu-Yang monopole for pure SU(2) Yang-Mills theory. 
Assuming that the function $f(r)=1+g(r)$ in equation (\ref{potential}) can be written as a power series defines, together with the requirement of
finite energy, the boundary conditions at infinity and at $r = 0$, namely: $g(0) = \pm 1$ and $g( \infty ) = \pm 1$. 
Besides the constant solutions $g(r) = 0$ and $g(r) = 1$, Wu and Yang were able to find, long ago,
a numerical solution with $g(0) = 0$ and  $g( \infty ) = -1$. 
 
In order to solve the classical equations of motion, we rederived the asymptotic series solutions, close to $r = 0$, named $A(r)$, 
and for large $r$, named $B(r)$. 
The first terms of the series solutions are given in equations (\ref{serie_A18}) and (\ref{serie_B9}), respectively. We have investigated the
two series and tried to build a solution of the equation of motion matching  $A(r)$ up to order $r^{18}$ and $B(r)$ up to order $1/r^{18}$. 
It turns out that the radius of convergence make such a type of solution impossible. Indeed, we found that increasing the order of the expansions
leaves almost unchanged the naive radius of convergence discussed at the end of section \ref{DSolSeries}.

In section \ref{DSolNum} numerical solutions of the equations of motion were computed and the behavior of their chromomagnetic fields discussed
at small and large distances. Our numerical procedure found two different solutions. The numerical solution with $g(0) = 0$ and  $g( \infty ) = -1$
discovered by Wu-Yang, which we named $g_{ext}(r)$, and a new solution $g_{int}(r)$. 

In what concerns $g_{ext}(r)$, we were able to provide an analytical description of its behavior at small distances in (\ref{WuYangNum_r0}). Further,
the classical energy associated with $g_{ext}(r)$ diverges due to its behavior at $r \sim 0$. On the other hand, the energy associated with
$g_{int}(r)$ is finite but one is forced to introduce a discontinuity on the first derivative of $f(r)$. Anyway, we have explored the conformal invariance of
the theory to set a scale for $g_{int}(r)$ assuming that its classical energy is a typical hadronic scale. 
It turns out that the discontinuity, assumed to give a measure of the monopole size, should happen at a length scale which is, at least, four times 
larger than the typical hadron size. Of course, if one requires that the discontinuity reproduces the typical hadronic size, then is
the monopole energy which is to large, at least by a factor of 8, than a typical hadronic mass. However, if one uses the maximum of the
energy density to define the monopole spatial extent, we get a size of $\sim 6.6$ fm or $\sim 3.9$ fm assuming a classical
energy of 1 GeV or 1.7 GeV, respectively, for the monopole.

In section \ref{SolDescontinuas} we explore solutions of the classical equation of motion which allow for
discontinuities on the first derivative. In particular, a Dirac delta function is added to equation (\ref{feq1}). As discussed, such a contribution can be
motivated by the presence of quarks. The introduction of a Dirac delta function $\delta (r - R_0)$ with a fixed $R_0$,
breaks explicitly invariance of the equations of motion under a scale transformation. 
Three different discontinuous monopoles configurations (\ref{monopolod1}) - (\ref{monopolod3}) were computed. If one uses a typical hadronic
scale to define the monopole energy, then, depending on the scale, $R_0 = 7.7$ fm or $R_0 = 2.4$ fm or larger. Again, this is quite a large
length scale. Probably, this means that the monopole configurations computed here play a minor role, if any, on low energy phenomenology.

Curiously, if one takes the critical energy density $\epsilon_c = 0.7$ GeV/fm$^3$ for the transition to the plasma phase \cite{plasma}
to set the scale, i.e. one puts 
$\epsilon_c = (\tilde{\mathcal{E}} / \lambda) / ( 4 \pi R^3 /3)$, 
where $\tilde{\mathcal{E}}$ is the dimensionless monopole energy, and take $R = 1$ fm, a typical hadronic length, then one gets a monopole
energy of 2.9 GeV. Using again the discontinuity on the first derivative as a measure of the monopole size, it follows that for the solution
$g_{int} (r)$, described in section \ref{DSolNum}, $\lambda = 2.26$ fm and $R_0 = 4.6$ fm. Remember that in section \ref{DSolNum},
$R_0$ was estimated to be 13.4 fm or 7.9 fm, depending on the choice of scale. 
For the solution $g_1(r)$ of section \ref{SolDescontinuas}, $\lambda = 1.16$ fm and $R_0 = 1.38$ fm, to be compared with 4.0 fm or 2.4 fm 
found in section \ref{SolDescontinuas}.
For the solution $g_2(r)$, $\lambda = 1.29$ fm and $R_0 = 2.6$ fm, to be compared with 7.7 fm or 4.5 fm found in 
section \ref{SolDescontinuas}.
For the solution $g_3(r)$, $\lambda = 1.41$ fm and $R_0 = 1.2$ fm, to be compared with 3.6 fm or 2.1 fm found in 
section \ref{SolDescontinuas}.
For the solution $g_4(r)$, $\lambda = 1.42$ fm and $R_0 = 2.4$ fm, to be compared with 6.9 fm or 4.1 fm found in 
section \ref{SolDescontinuas}.
This suggests that these configurations may play an important role in the confining region and in the transition to the plasma phase. 
Furthermore, the so-called discontinuous solutions, with sizes in the range $1.2 - 2.6$ fm, are considerable shorter than $g_{int}(r)$. 
The difference is certainly due to the Dirac-delta term introduced in the equation and can be seen as a clear sign of the role of the breaking
of conformal invariance due to the coupling to the quark fields.
Hopefully, one would be able to investigate possible signatures for identifying the monopole configurations
in the transition to the plasma phase,  in ongoing and future experiments to probe this state of matter with high-energy
heavy ion collisions.

\section*{Acknowledgments}

This work was partially supported by Capes/FCT project 183/07. TF and BVC acknowledge partial
financial support from FAPESP and CNPq.
JAOM acknowledges financial support from CAPES under contract number BEX4092/08-2.



\begin{thebibliography}{99}

\bibitem{actor} A. Actor, Rev. Mod. Phys. \textbf{51},  461-525 (1979).

\bibitem{dyon} B. Julia, A. Zee, Phys. Rev. \textbf{D11}, 2227-2232 (1975).

\bibitem{cp} A. Actor, Z.Phys. \textbf{C6}, 223-234 (1980).

\bibitem{popov} See, for example, A. D. Popov, J. Math. Phys. \textbf{46}, 073506 (2005) and references their in.

\bibitem{WuYang} T.T. Wu, C.N. Yang, 
in \textit{Properties of Matter Under Unusual Conditions}, edited by
  H. Mark and S. Fernbach (Interscience, New York, 1968).

\bibitem{MSW} J. Moody, A. Shapere, F. Wilczek, Phys. Rev. Lett. \textbf{56}, 893 (1986).

\bibitem{QGP} J.-P. Blaizot, E. Iancu, Phys. Rep. \textbf{359}, 355-528 (2001).

\bibitem{stuler} R. L. Stuller, Phys. Rev. \textbf{D22}, 2510 (1980).


\bibitem{AD} Z.F. Ezawa, A. Iwazaki, Phys.Rev. \textbf{D25},  2681 (1982).

\bibitem{holdom} B. Holdom, Phys. Rev. \textbf{D79}, 085013 (2009). 

\bibitem{brodsky} G. F. de Teramond, S. J. Brodsky, Phys. Rev. Lett. \textbf{102}, 081601 (2009); and references therein. 

\bibitem{chen} Y. Chen et al, Phys.Rev. \textbf{D73}, 014516 (2006).

\bibitem{plasma} F.Karsch, E.Laermann, {\tt hep-lat/0305025}.



\end{thebibliography}
\end{document}